\def\kms{\ifmmode{\hbox{km~s}^{-1}}\else{km~s$^{-1}$}\fi}
\documentstyle[12pt,aaspp4]{article}
\tighten
\begin{document}

\title{Hubble Space Telescope Planetary Camera Images of NGC~1316
(Fornax A) \footnote[1]{Based on observations with the NASA/ESA {\it
Hubble Space Telescope}, obtained at the Space Telescope Science
Institute, which is operated by AURA, Inc., under NASA Contract No. NAS
5-26555.}}

\author{Edward J. Shaya, Daniel M. Dowling \& Douglas G. Currie}
\affil{Department of Physics, University of Maryland, College Park,
Maryland
20742. \break E-mail: shaya@img.umd.edu, dowling@img.umd.edu,
currie@img.umd.edu}

\author{S. M. Faber}
\affil {UCO/Lick Observatories, Board of Studies in Astronomy and
Astrophysics, University of California, Santa Cruz, California 95064.
\break E-mail: faber@lick.ucsc.edu}

\author{Edward A. Ajhar \& Tod R. Lauer} 
\affil{Kitt Peak National Observatory, National Optical Astronomy
Observatories\footnote[2]{The National Optical
Astronomy Observatories are operated by the Association of Universities
for Research in Astronomy, Inc. (AURA) under cooperative agreement with
the National Science Foundation.}, P.O. Box 26732,
Tucson, Arizona 85726. \break E-mail: ajhar@mars.tuc.noao.edu, lauer@noao.edu} 

\author{Edward J. Groth}
\affil{Physics Department, Princeton University, Princeton, New Jersey 08544.
\break E-mail: groth@pupgg.princeton.edu}

\author{Carl J. Grillmair}
\affil{UCO/Lick Observatories, University of California, Santa Cruz,
California 95064. \break E-mail: carl@lick.ucsc.edu} 

\author{C. Roger Lynds \& Earl J. O'Neil, Jr.}
\affil{Kitt Peak National Observatory, National Optical Astronomy 
Observatories$^2$, P.O. Box 26732, 
Tucson, Arizona 85726. \break E-mail: rlynds@noao.edu, eoneil@noao.edu}

\begin{abstract} 

We present HST Planetary Camera V and I~band images of the central
region of the peculiar giant elliptical galaxy NGC~1316.  These high
resolution images reveal that the central surface brightness rises
sharply to 12.1 mag arcsec$^{-2}$\ in the I~band and 13.5 mag
arcsec$^{-2}$\ in the V~band.  The inner profile is well fit by a
nonisothermal core model with a core radius of $0\farcs41\pm0\farcs02$\
(34 pc).  At an assumed distance of  16.9~Mpc,  the deprojected
luminosity density reaches $\sim 2.0 \times 10^3 L_{\sun}$ pc$^{-3}$.  

Outside the inner two or three arcseconds, a constant mass-to-light
ratio  of $\sim 2.2 \pm 0.2$ is found to fit the observed line width
measurements.   The line width measurements of the center indicate the
existence of either a central dark object of mass $2 \times 10^9
M_{\sun}$, an increase in the stellar mass-to-light ratio by at least a
factor of two for the inner few arcseconds, or perhaps increasing radial
orbit anisotropy towards the center.  The mass-to-light ratio run in the
center of NGC~1316 resembles that of many other giant ellipticals, some
of which are known from other evidence to harbor central massive dark
objects (MDO's).

The $V - I$ color of unreddened regions is found to be uniform at $1.55
\pm 0.10$. The profile does not get significantly bluer near the center
as might be expected in a merger, except for perhaps the innermost
0\farcs1.  Fits to the unextinguished regions at the center raise
concerns that the  possible UV-bright point source reported by Fabbiano
{\it et al.} (1995) is partially or fully explained by  dust clouds
wrapped around the central line of sight. The smoothness of the
underlying stellar distribution allows analysis of the 3-dimensional
distribution of the dust. We use two observables; the color excess and
the ratio of observed V~band flux to that of a symmetric smooth fit. 
The maximum optical depth is only A$_V \sim 1.5$, and only complexes on
the near side of the galaxy are detected. 

We also examine twenty globular clusters associated with NGC~1316 and
report their brightnesses, colors, and limits on tidal radii. The
brightest cluster has a luminosity of $9.9 \times 10^6 L_{\sun}$ ($M_V =
-12.7$), and the faintest detectable cluster has a luminosity of
$2.4 \times 10^5 L_{\sun}$ ($M_V = -8.6$). The globular clusters
are just barely resolved, but their core radii are too small to be
measured. The tidal radii in this region appear to be $\le$ 35~pc. 
Although this galaxy seems to have undergone a substantial merger in the
recent past, young globular clusters are not detected.

\end{abstract}


\section{Introduction}

NGC~1316, in the Fornax Cluster, is a giant Morgan type D elliptical
with a NW-SE dust lane.  It has a strong double-lobed radio source,
Fornax A, and an unresolved radio core (Geldzahler  \& Fomalont 1978,
1984).  Fabbiano {\it et al.} (1995) report a possible unresolved bright
point source at $\lambda1750$\AA\ at the center.    The outer region has
shells, loops, ripples and possibly a tidal tail (Schweizer 1980),
evidence of a recent merger event.  The brightness profile at the center
rises quite sharply.  These assorted observations make NGC~1316 a likely
candidate for hosting an active supermassive black hole.  Unfortunately,
the inner region is partially obscured by the  dust lane, and this
complicates the  determinations of the brightness profile at the center. 
We report here on information at additional wavelengths, the V and
I~bands, that help to sort out some of the extinction difficulties and
determine the profile in the inner few arcseconds.

Schweizer (1981) found a central velocity dispersion of 248 km
sec$^{-1}$ and a core radius, $r_c$, of $0\farcs6 \pm 0\farcs2$ after
correction for the $FWHM = 0\farcs90$ seeing.  He derived a
mass-to-light ratio in the V~band of 2.5 (correcting to a distance of 16.9
Mpc).  This low value would appear to be contrary to the
idea that a supermassive object resides at the center.     In \S 3, the
core size is found to be $\sim 0\farcs41$.  Hydrostatic models,
presented in \S 4, indicate a mass-to-light ratio of $2.2 \pm 0.2$
outside of the inner few arcseconds, but at the center, additional dark mass 
appears to be required.

In \S 5, estimates are made of the amount of extinction present and a rough
determination is attempted of where along the line of sight the
extinction resides.  We arrive at estimates of extinction at the center
and re-evaluate the photometry of the central region.

We examine, in \S 6, the brightest globular clusters in this galaxy to
determine whether the merger resulted in the formation of young clusters
similar to those in NGC~1275 (Holtzman {\it et al.} 1992), NGC~3597,
NGC~6052 (Holtzman {\it et al.} 1996), NGC~7252 (Whitmore {\it et al.}
1993), and NGC~4038/4039 (Whitmore \& Schweizer  1995).  No such bright
young clusters are detected.  Limits are placed on the cluster tidal radii
assuming King (1966) models.

Although NGC~1316 lies at the outskirts of the Fornax Cluster, it has
been established to be a member of Fornax by both the surface brightness
fluctuation method (Tonry 1996, personal communication) and its
planetary nebulae brightness distribution (McMillan {\it et al} 1993).
To obtain the distance to the Fornax Cluster, we average the differences
between distance moduli measured for the Virgo Cluster  and the Fornax
Cluster from: the planetary nebulae brightness distribution,
$\mu_{Fornax} - \mu_{Virgo} = 0.24\pm0.10$ (McMillan {\it et al} 1993); 
the $D_n-\sigma$ relation, $\mu_{Fornax} - \mu_{Virgo} = 0.25\pm0.31$
(Faber {\it et al.} 1989); and the surface brightness fluctuation
method, $\mu_{Fornax} - \mu_{Virgo} = 0.08\pm0.11$ (Tonry 1991).  The
average $\mu_{Fornax} - \mu_{Virgo} = 0.176$ can be added to the 
distance determination of the Cepheids in the  Virgo Cluster of 30.96
(d$_{Virgo}$ = 15.6 Mpc) (Freedman {\it et al.} 1996) to derive 
$\mu_{Fornax} = 31.14\pm0.40$ (16.9 Mpc).  This distance is in
relatively good agreement with peculiar flow model calculations (Tully,
Shaya, \& Pierce 1992) of 19 Mpc. The redshift of 1635 km s$^{-1}$ is
substantially increased by reflection of our own peculiar velocity
toward the Virgo Cluster direction.  Most of the previous studies of
this galaxy assumed distances that are too large because the distances
were  based  on unperturbed Hubble flow and a lower value for $H_0$.  At
the closer distance, the galaxy fits well to the Faber \& Jackson (1976)
relation between central velocity dispersion and absolute brightness. 
However, the central surface brightness is 3 mag arcsec$^{-2}$ brighter
than expected for a galaxy of total absolute magnitude of $M_B = -21.80$
(Sandage \& Tammann), according to the relations of Kormendy (1987) and
Faber {\it et al.} (1996).

\section{Observations and Reduction}

As part of a WF/PC-1 GTO investigation of galaxies with interesting
nuclei,  optical images were taken with the HST Planetary Camera (PC) on
1991 July 2.  The telescope was affected by spherical aberration during
this period.  The PC is composed of four 800 x 800 pixel CCDs. The
pixels subtend 0\farcs0437 and the total field of view is $70\arcsec$.
The center of NGC~1316 appears in PC6. Data were obtained with the F555W
and F785LP filters, corresponding roughly to Johnson V and I. 
Two 260 second images were taken in each filter.   Observations in both
filters of a bright star, exposed on the same day as the galaxy
observations, were used as the point spread functions (PSF). The F785LP
exposure time was 0.26 s and the F555W exposure was 0.11 s.  The guiding
mode during all exposures was coarse track.  The rms jitter from
telescope guiding errors, according to the log of guide star motions
obtained from the OMS branch of STScI, was $\approx 30$ mas  in each
N~1316 image. When the PSFs are convolved with gaussians with rms of 30
mas their radial profiles become similar to that of the bright Galactic
star in the galaxy image.  This is consistent with there being
negligible jitter in the short exposure PSF images.          

Images were processed according to the procedure outlined by Lauer
(1989). Shifts between images could be accurately determined from the
position of the nucleus and several bright knots scattered through the
images, which we identify as globular clusters.  Individual images for a
single filter shifted by $< 0.1$ pixel in position, so we simply
co-added these. There was a shift between the V and I filters by 9
columns and one row.  The galaxy was detected to the edges of all four
chips.  We present deconvolved images  from the PC
(Figure~\ref{fig1}) with wide logarithmic stretch.  The dust lane is
clearly evident in the F555W filter, but barely noticeable in the F785LP
image.  A few weakly visible features due to dust grains in the
optical path of the WFPC-1 camera were patched by interpolation.

When the V and I summed images are divided by one another there is a
very  faint, thin  ellipse  discernable with a semimajor axis of 130
pixels.  This feature occurs at locations where the count level is
256 ADU.  It is almost certainly an artifact of the  dropouts of A-to-D
codes at the 256-bit transition (Lauer 1989) of the A-to-D converter in
the WFPC-1 camera. This feature is easily ignored.
                
Sixty iterations of the Lucy-Richardson deconvolution procedure (Lucy
1974)  were applied to the inner 512 by 512 pixel region to partially
correct for the spherical aberration introduced by the HST optics.  When
more than sixty iterations were applied, the noise was amplified to
undesirable levels.  Since the signal-to-noise is high ($\approx 130$)
in the central region, we also deconvolved using a Weiner filter
modified to further reduce noise. The two methods agreed exceedingly
well (average rms deviation was $\approx 1\%$  within 50 pixels of the
nucleus).   

Recalibration of the zero-points in the conversion to Johnson V and I~bands was
accomplished by referring to ground-based aperture photometry (Poulain
1988).  We used Bessel's (1979) transfer equation $(V-I)_C =
0.778(V-I)_J$ to convert the ground based Cousins I to Johnson I. 
Comparisons were made to apertures of 22\farcs87  for V  and
31\farcs21 for I.  Our calibrations give 23.11 and 21.60 for V and I
band zero points; the WF/PC Science Verification Report (WF/PC Team 1992)
values  for PC6, based on individual stars, are 23.05 and 21.55
respectively. 

\section{Brightness Profile}
         
The dust lane presents a problem in attempting to measure the brightness
profile in the central few arcseconds.   Penereiro {\it et al.} (1994)
measured the Gunn-Thuan $g-r$ color profile of this galaxy and found
that it is quite constant from 7\arcsec\ to 25\arcsec.  Interior to
7\arcsec\  there is a rapid rise in $g-r$ of 0.2 mag.  With HST
resolution we can clearly see that this rise is due entirely to
the dust lane.  Apart from regions belonging to the dust lane and the
central few pixels, the $V-I$ color in the PC chip is uniform at
$1.55\pm0.10$  (Figure~\ref{bigvmi}). Much of the quoted uncertainty in
the color can be ascribed to the precision of the flat fields. 

We can make an estimate of the amount of dust extinction for the central
pixels.  The  contours of the $V-I$ image (Figure~\ref{vmicr}) indicate
that $V - I < 1.8$ for the central 5 pixels in radius.  There are two
possible situations by which dust extinction can result in low color
excess;  either the extinction is so complete that almost none of the
light from the center or behind it is transmitted, or the extinction is
very light.  There is also the possibility that the bluing trend toward
the center is due to blue stars near the nucleus.  The key to deciding
between these possibilities is to note that the dust is definitely not
distributed elliptically like the light.  Since the I~band light is
distributed fairly smoothly in elliptical isophotes near the center and
the I~band and V~band give consistent centroids that are aligned with
elliptical contours at larger radii, we conclude that the center has
only a small amount of extinction, with upper limit values of $A_V <
0.4$ mag and $A_I < 0.2$ mag. 

To compute brightness profiles, we solved for the intensity and ellipse
shape which gave the minimum square deviation from median intensity for
each chosen semimajor axis.  Two gaussians were fit, one along columns
and one along rows, to the inner 7 pixels in radius to determine the
centers.  The centers of the ellipses were not varied, but the
eccentricity and orientation did vary.  Dusty regions were excluded from
the calculation by ignoring pixels with $V-I > 1.6$. 
Figure~\ref{magprof} and Table 1 present brightness as a function of
semimajor axis for the raw and deconvolved images of both bands.   The
large swing in position angle of the major axis within 0\farcs5 is
probably just an artifact of dust obscuration that was not fully
removed. 



The central pixel of the $V-I$ map before deconvolution is bluer by
about $\approx 0.1$~mag, and by $\approx 0.2$~mag in the deconvolved
image (Fig. 3).  The blue region appears to be only about 3 pixels in
radius. However, this can be ascribed entirely to the higher resolution
at shorter wavelengths.  When we convolve the deconvolved F555W image
with the F785LP PSF the central pixel does not appear to be bluer than
the rest of the image.  In other words, the deconvolved V profile is also a
deconvolution solution for the I profile within the errors. Thus, the
inner 0\farcs04 of this galaxy may be bluer by $\approx 0.2$~mag, but we do
not have sufficient information at this scale to make that claim.  

We now fit the deconvolved I~band surface brightness profile.  It should be
recognized that the deconvolution process does not produce an image of
infinite resolution.  The central pixel may still be underestimated by 
$\approx 0.4$~mag due to the finite resolution and reddening.  We use the
functional form 

\begin{equation}  \Sigma(r) = 2^\eta \Sigma_b  \Bigl[1+\Bigl({r \over
r_b}	\Bigr)^\alpha\Bigr]^{-\eta},  \label{eq_nukeslaw} \end{equation}
This form is the $\gamma = 0$ case of the function used by Byun {\it et
al.} (1996) and  Lauer {\it et al.} 1996) to represent surface
brightness profiles for several elliptical galaxies observed with {\it
HST}.  In the general form, $r_b$ is considered to be the breakpoint
from an outer slope to an inner slope, $\gamma$.  The $r_b$ parameter
can be considered to be a core size.   We determined least-squares
values of the parameters for the deconvolved I~band profile. Best fit is
found with  parameters  $r_b=0\farcs41\pm0.02$, $\alpha=1.16\pm0.02$,
$\eta=1.00\pm0.02$, and $\Sigma_b =  12.88 \pm 0.06$
I~mag~arcsec$^{-2}$, where the errors are determined by holding the
other parameters fixed. The fit is  shown as the top solid line in
Figure~\ref{magprof}.  We also plot as a solid line on this figure the
I~band fit increased by a constant 1.55 which fits the V~band profile
quite well, except, perhaps, for the central pixel or two.  Thus the
$\Sigma_b$ value in V is 14.44 mag~arcsec$^{-2}$ which, at 16.9 Mpc is
equivalent to $5.37\pm0.30 \times 10^4  L_{\sun}$ pc$^{-2}$.  


\subsection{Deprojected Density Profile}

We use the Abel deprojection scheme (Bracewell 1978) to deduce the three
dimensional density profile from the projected major axis profile.  When
the swings in position angle are small, this procedure implicitly
assumes that the figure of the galaxy is an oblate spheroid (solid line
in Figure~\ref{lumden}).  The procedure was repeated using the profile
along the minor axis, which is equivalent to assuming that the galaxy is
prolate (dashed line).  The galaxy has substantial rotation about the
minor axis, so it is
unlikely to be fully prolate, but by calculating the density run for
both prolate and oblate we bracket  a wide range of spheroidal shapes
and demonstrate that the range of variation in densities is small. The
three dimensional luminosity density falls off as $r^{-2.0}$ for radii
$r > 1\arcsec$, and the central density reaches $\ge 2000 L_{\sun}$
pc$^{-3}$.  This range of densities is now becoming quite common in HST
observations of nearby  galaxies (Lauer {\it et al.}  1996).

The deprojected density profile is also well fit by
Equation~\ref{eq_nukeslaw}.  The values of the parameters of the best fit to
the oblate shape are: $r_b = 0\farcs265\pm0.015$, $\alpha=1.80\pm0.03$,
$\eta=1.13\pm0.03$, and $\Sigma_b = 654 \pm 25 L_{\sun}$ pc$^{-3}$. 
It is noteworthy that based both on the projected and deprojected
profiles, NGC~1316 has one of the {\it flattest} interior density cusps,
compared to the sample of Lauer {\it et al.} (1996).

\section{Mass-to-Light Ratios}

Schweizer (1981) measured a line of sight central velocity dispersion of
$\sigma_v = 248 \pm 6$ km sec$^{-1}$ that refers to a region $6\farcs7
\times 1\farcs1$.  Jenkins \& Scheuer (1980) measured $\sigma_v=262 \pm
6$ km sec$^{-1}$ in a region $2\farcs0 \times 2\farcs5$.  Based on these
observations, an assumed core radius of $R_{hb}=0\farcs6\pm0\farcs2$, a
central surface brightness of $\Sigma_{V,0} = 13.3\pm0.3$
mag~arcsec$^{-2}$, and the standard equation for the mass-to-light ratio
in a galactic core,
 
\begin{equation} 
M/L = {9 \sigma^2 \over 2\pi G \Sigma_0 R_{hb}},\label{eq-ml}  
\end{equation} 

\noindent where $R_{hb}$ is the half-maximum brightness radius,
Schweizer quoted a V~band mass-to-light ratio for the inner core of
$1.3\pm0.2$ (H$_o=50$).   When converted to our adopted distance this
corresponds to a mass-to-light ratio of 2.5.  However, Equation 1 is
appropriate only for an isothermal core.    For example, a brightness
profile that follows the de Vaucouleurs' law will have its 
mass-to-light ratio underestimated by a factor of 7 (Richstone \&
Tremaine 1986).  There is a more general procedure to determine
mass-to-light ratios which we shall now describe.

The light profile is distinctly different from that of an isothermal
distribution. However, the velocity dispersion profile is, as we shall see,
fairly constant. Either a varying mass-to-light ratio or a strongly
anisotropic velocity dispersion is required to explain the
shape of the light profile.   We take an approach toward constraining
the mass-to-light ratio by modeling the run of kinematic pressure with
radius assuming an isotropic velocity dispersion.

We can combine the new light profile with previously published velocity
profiles to determine the distribution of total mass  and mass-to-light
ratios in the central 1 kpc. Bosma {\it et al.} (1985) measured the run
of velocities with radius.  They found the velocity dispersion falls
linearly from $232\pm13$~\kms\ at the center to about $150\pm50$\ \kms\
at 100\arcsec.  The rotational velocity rises linearly from  $0\pm5$\
\kms\ at the center to about $150\pm 50$\ \kms\ at 100\arcsec.  Their slit
was aligned along pa = $60\arcdeg$, which is very close to our estimates of
the major axis position angle (Table 1). 

The expected run of velocity dispersion $\sigma$ required for
hydrostatic equilibrium can be calculated from the light distribution
and some assumed profile of the mass-to-light ratio by calculating the 
pressure at each radius arising from the material outside that radius.  
NGC~1316 is substantially flattened, so the potential for oblate
spheroidal shells of eccentricity $e$ is used (Binney \& Tremaine 1987,
P52):

\begin{equation} 
\Phi(a') = -{G \delta M \over ea} \arcsin\Bigl({ea \over a' }. \Bigr) 
\end{equation} 

\noindent Here the major-axis radius of the shell is $a$, and the mass
of the shell is  $\delta M = 4 \pi \rho a^2 (1-e^2)^{1/2} \delta a$.  
The acceleration contributed by each shell to a point on the equatorial
plane at radius $a'$ is: 

\begin{equation} 
g(a') = - {\partial \Phi \over \partial a'} = 
		{-G \delta M \over a'^2[1-({ea \over a'})^2]^{1/2}}. 
\end{equation} 

\noindent The acceleration at $a'$ is reduced by rotation in the amount 
$V_{rot}^2(a')/a'$.   

For the overpressure exerted on our outermost radius, we simply assume a
power law exponent of $\beta=2$ for the density beyond the outer radius.
We assume a constant rotational velocity of 150 \kms\ outside the last
measured data point.  Integrating density times gravitational
acceleration from outer radius $r_m$ to infinity leads to the radial
presure at $r_m$:

\begin{equation} 
P_r(r_m) = {2 \pi G \rho_m^2 r_m^2 \over {\beta^2 - 1}}  + 
	{G M(< r_m) \rho_m \over {r_m(\beta + 1)}} - \rho_m V_{rot}^2 / \beta. 
\end{equation}

\noindent Working inward, one shell at a time, the additional weight is
calculated and pressure equilibrium establishes the required radial
velocity dispersion, $\sigma_r$, of each shell.  The eccentricities of
the projected image were used to approximate the true cross-sectional
eccentricities.   Figure~\ref{vel_r} presents $\sigma_r(r)$ assuming a
constant V~band mass-to-light ratio of 2.2.  If no rotation is assumed
(dashed line), the expected velocity dispersion rises slowly with
radius. The inclusion of rotation (solid line) reduces some of the
pressure and we find $\sigma_r$ turns over at several hundred pc.  

To compare a model with observations along a slit, it is necessary to
integrate $\sigma_r^2$ through the line of sight and weight by the
density at each position along the path.  For the observations near the
center, one must also integrate over all lines of sight within the
aperture.  Figure~\ref{vels_R} shows the resultant projected velocity
dispersion along a slit assuming the dispersion is isotropic.  The fit
to the data points of Bosma {\it et al.} (1985) beyond 2\arcsec\ is
quite good for a model with a constant mass-to-light ratio of 2.2 (solid
line). The {\it central pixel} data points of Jenkins \& Scheuer (1980)
(square), Schweizer (1981) (triangle) and Bosma {\it et al.} (1985)
(plus) are shown on Figure~\ref{vela_R} for equivalent circular
apertures.  They all lie substantially {\it above} the curve established
by data at larger radii.  The effective apertures in each case were
limited by the quoted seeing.  The Jenkins \& Scheuer dispersion
measures run about 10\% higher than the Bosma {\it et al.} measurements
at all radii on this galaxy, and so there may need to be a slight
adjustment to one or the other. 
                  
One obvious possibility that would raise the model velocities at $\sim
1\arcsec$  to better match the observations is the addition of a massive
central point mass.  The dotted lines in Figures~\ref{vel_r},
\ref{vels_R}, and \ref{vela_R} show model velocity runs when point
masses of $3 \times 10^8, 1 \times 10^9,$ and $3 \times 10^9 M_{\sun}$
are added to the above model with mass-to-light ratio of 2.2.  It is
seen that a $2\times10^9M_{\sun}$ dark compact object would give a
reasonable fit to the central velocity dispersion measurements. Of
course, there is an obvious prediction of substantially higher
velocities at smaller radii, but we are unaware of published data with
better spatial resolution.

Another way to explain the excess in observed velocity dispersion at the
center is to allow  stellar mass-to-light to vary.   An easy way to
proceed becomes apparent when one realizes that the total velocities,
from 1\arcsec\ to 100\arcsec, are not significantly different from a
constant value of 240 \kms\ if one takes into account that much of the
dropoff in velocity dispersion is compensated by rotation. Thus, we may
use the well known isothermal density distributions to arrive at a good
approximation.  At large radii, the luminosity density falls as $1/r^2$,
as does an isothermal distribution.  But in isothermal distributions,
the profile steepens below an inverse square law, just outside the core
radius.  Since the density distribution of NGC~1316 does not follow
this, significant mass-to-light variation is required to obtain constant
$\sigma$.  If one simply divides the luminosity density into the family
of King isothermal profiles that result in $\sigma_v=240$, one gets the
family of mass-to-light runs shown in Figure~\ref{emol}, where the
choices of core radii were 0\farcs2, 0\farcs4, 0\farcs8, and 1\farcs6,
corresponding to 16.8, 33.6, 67.2, and 134 pc.  

Certain cases can be excluded as being unreasonable.  The largest of
these core radii can be excluded because it has mass-to-light rising, at
first, as one moves inward and then dropping below unity near the
center.  Given the lack of color gradients, this behavior does not seem
reasonable.  The next largest core radius is also unattractive because it has
a mass-to-light bump at a random radius in the galaxy; one would be hard
pressed to explain such behavior.  The other two cases are both fairly
acceptable.  They contain higher mass-to-light with decreasing radius,
except at  radii smaller than the resolution limit.  The
conclusion is that the stellar mass-to-light ratio is $2.2 \pm 0.2$ at large
radii and rises rapidly within a 2\arcsec\ to values of at least 5.  A
massive dark compact object would not then be required.  

NGC~1316 is typical of many spheroidal galaxies in showing a modest rise
of $10-20$\% in central velocity dispersion at groundbased resolution. 
It was this behavior in M87 that inspired the suggestion that it
contained a central dark object of mass $3 \times 10^9 M_{\sun}$
(Sargent {\it et al.} 1978).  This argument subsequently was disputed
when it was realized that velocity anisotropies could also match the
data (Binney \& Mamon 1982).  This same explanation can clearly be
invoked for NGC~1316 as well.  However, subsequent observations of
promising dark mass candidates at high resolution (including M87) have
invariably revealed further increases in central velocity dispersion
(and rotation) that are fully consistent with actual central dark
masses.  The accumulating number of such cases (see Kormendy \&
Richstone 1995 for a review) plus the known non-thermal nuclear
activitiy in NGC~1316 suggest that it, too, contains a dark mass.  High
resolution velocity dispersion measurements with HST, planned for late
1996 or early 1997 with the Faint Object Spectrometer, should test this
possibility.
                                 
Finally, a few words on the global mass-to-light ratio, $M/L_V = 2.2 \pm
0.2$ or $M/L_R$ = 1.0,  assuming $V - R$ = 0.7. This is remarkably low
compared to normal ellipticals of comparable luminosity, which have an
average $M/L_R \sim 4.0$ (van der Marel 1991, adjusted to $H_0 = 80$). 
This low value might be associated with star formation triggered by the
recent merger.  More insight might come from comparing global parameters
such as radius, surface birghtness, dispersion, and colors versus those
of normal ellipticals of comparable mass. 

\subsection {Core Collapse}

One might be persuaded that a black hole could have formed via
gravothermal collapse in the center if the collapse time of the core
were shorter than a Hubble time.  Assuming an isotropic velocity
dispersion and stars of about one solar mass, the relaxation time is
(Binney \& Tremaine 1987, P. 514),

\begin{equation} t_r = 3.7 \times  10^{12} yr  \Bigl({\sigma \over
{250~{\rm km~s^{-1}}}}\Bigr)^3  \Bigl({ {4 \times 10^3~M_{\sun} {\rm
pc}^{-3}} \over{\rho} }\Bigr) \ln\Bigl(0.4{N\over5.2e8}\Bigr)^{-1}. 
\end{equation} Fokker-Planck based simulations have shown that
single-mass systems collapse after $\approx 15$ relaxation times of the
core (Cohn 1980)  and multi-mass models can collapse in as short as 2
relaxation times (Inagaki \& Saslaw 1985).  Thus the core collapse time
is $10^{13}$ yr.  If a supermassive black hole developed in
NGC~1316, it must have formed by gaseous dissipation.

\section {Color Analysis and Dust}

\subsection{The Smoothed Fit, $V_{fit}$} 

We also attempted to fit the deconvolved V~band image, ignoring regions
of dust obscuration.  The purpose is to measure the amount of light
removed along the lines of sight by the dust clouds, independent of the
reddening determination.  In the inner regions of this galaxy, a large
fraction of each isophote includes dust clouds, rendering the
eccentricity and position angles of ellipse fitting uncertain.  We
therefore refit with a more elaborate procedure that permits control of
the degree of fit and requires a minimum amount of interpolation. The
brigtness profile  was remapped to make the isophotes circular, and the
resulting image was fit along circles by sine series.  Each sine term
had an independent amplitude and reference angle.  Pixels with $V-I >
1.6$ were masked out.  The degree of the fit was varied according to the
fraction of pixels unmasked at each radius. For the inner 25 pixels, the
fit was only to a constant surface brightness.  From 26 pixels to 40
pixels, a constant and the first sine term were used.  Beyond 40 pixels,
the fit contained a constant and two sine terms.  Upon completion of
this fitting, it was apparent that there were dust regions of low color
excess. Therefore, the mask was augmented by pixels that were below the
fit by 50 ADU or more, and the process was repeated. This process
smoothly filled in regions with evident dust absorption.   The
difference between the observed V~band image, $V_{obs}$, and the fit,
$V_{fit}$, is shown in Figure~\ref{v60fitm}.
 
There is a region of negative values in the difference in the  inner 25
pixels but these are all less than 10\% of the flux. A cut down the
major axis of this fit gives a very similar profile as the ellipse
fitting of \S~3 (Figure~\ref{v60profs}).  This gives us additional
confidence that our fitting of the brightness profiles correctly gives
the unextinguished profiles to within about 0.2 mag in the I~band  in
the inner arcseconds. This central region has an incomplete loop of
absorption going around the central 5 -- 10 pixels.  In the I-band
image, the absorption band is just barely discernible.    However, at
$\lambda 1750$\AA\ the absorption is an order of magnitude greater, and
one should be wary, as Fabbiano {\it et al.} (1995) are, that the UV
bright point source reported by them may be merely an artifact of the details
in the dust distribution. 

Figure~\ref{radio} presents an overlay of a radio map at 1.5 GHz
produced by Geldzahler \& Fomalont (1984) on the $V - I$ grayscale
image. We note that the SE jet terminates at one of the thick clouds. 
This indicates that the SE jet is pointing toward us since we find that
the detected dust clouds are in front of the midplane.  Presumably the gas or
dust has sufficient density to stop the charged particles responsible
for the radio emission. The NW jet extends farther, apparently missing
any region of high gas and dust density.  If this geometry is correct,
it suggests that the jet is not highly relativistic, as the side away
from us is visible.

\subsection{Color Excess -  ${V_{obs} \over V_{fit}}$ Relation}

The smoothness of the underlying elliptical galaxy allows  measurements
of the dust distribution to be made that are rarely possible. We compare
the color excess, $E(V-I)$, to the ratio of observed to flux to the fit
flux of the previous section,
$V_{obs} \over V_{fit}$.  A simple starting model assumes that each line
of sight passes through a single discrete cloud within the galaxy that
dominates the extinction.  We define $V_{front}$ to be the flux
emanating from stars in the galaxy between the observer and the single
cloud along the line of sight.  We define $V_{back}$ to be the flux
emanating from stars directly behind the cloud as seen by the observer,
and similarly for $I_{front}$ and $I_{back}$.  We assume constant
intrinsic color along the line of sight.  The ratio of the two fluxes is
designated $s$:

\begin{equation} 
s \equiv {{V}_{front}\over{V}_{back} }={{I}_{front}\over{I}_{back} }.
\end{equation}

\noindent The expressions for color excess and $V_{obs} \over V_{fit}$
are then,	

\begin{equation}
E(V-I)=-2.5\log\Bigl[{s+10^{-0.4A_V}\over s+10^{-0.4A_I}}\Bigr]
\end{equation}
and
\begin{equation}
{V_{obs} \over V_{fit}} = { s + 10^{-0.4A_V} \over {1+s} }.
\end{equation}
For Figure~\ref{avs}, we also define the `depth of cloud', 
\begin{equation}
{s_2={{V}_{front}\over{V}_{total}}}={{s}\over{s+1}}.
\end{equation}

Planck functions and stellar spectra have been multiplied by standard
reddening curves and then multiplied by HST/WFPC-1 throughput curves to
derive the relation $A_V = 2.0A_I$ (Shaya {\it et al.} 1994).  Curves of
constant $s_2$ (dashed lines) and constant $A_V$ (solid lines) are
plotted in the $E(V-I)$ versus ${V_{obs} \over V_{fit}}$  plane in
Figure~\ref{avs} along with the data for all pixels within 7\arcsec\ but
outside of 1\arcsec\ from the center.  

Most of the data points lie in a region in Figure~\ref{avs} that should
be impossible for dust with $A_V = 2.0 A_I$.  At first, we suspected
that this galaxy may have unusual dust optical properties, and found
that $A_V = 1.6 A_I$ would fit quite well.  However, if the extinction
has substantial variations on smaller scales than the resolution, there
could be a small shift of the data points on this type of plot.  Light
scattered into the line of sight by resolution blurring has a greater
effect on the $V_{obs}/V_{fit}$ than on the color. We suspected that a
shift of this nature could be caused by incomplete deconvolution.  We
ran simulation tests in which we started with a constant intensity image
and then decreased the intensity in disks to simulate variously sized
disks of dust with a variety of optical depths.  The image was convolved
using the HST PSF, noise was added, and then the image was deconvolved
using 60 Lucy-Richardson iterations.  We found the distribution of
points tended to move in the correct direction.  A few examples of the
test results are shown in Figure~\ref{avmodels}.  The dust model that
most resembled the observations had $A_V = 1.0$, radius of cloud = 10
pixels, and $s_2$  = 0.0 (completely in front).

None of the  points in Figure~\ref{avs} lie in the region of very high
extinction.  High extinction would appear as a large drop in surface
brightness but with little reddening because most of the observed light
arises in front of the absorbing region.  If there were much dust at the
midplane with large values of extinction, then one would expect to find
many pixels along the dust lane with about 50\% of the light missing
($A_V = 0.7$) and $E(V-I) < 0.2$, but {\it no redder}.   However, values
near this range occur for only two pixels.  The vast majority of pixels
show large color excess and relatively little absorption. This is the
signature of low extinction and low depth along the line of sight.

In addition, near the centers of each of the prominent complexes there
are regions with nearly 80\% of the total light removed and very large
reddening ($E(B-V) \sim 0.6$).  This suggests that all of the prominent
complexes are in front of at least 80\% of the projected starlight (30\%
of the light in front of the midplane).  And, the extinction of the
clouds all have $A_V < 1.5$.   That the extinctions are typically so
small is actually quite obvious from the I~band image alone because the
dust, in that image (Figure 1b) is fairly difficult to perceive.  The
dust is probably distributed in a patchy ring.  We do not detect the
dust on the far side because the fraction of light removed from the line
of sight is too small.  It is even possible for the clouds to be
randomly  distributed within a plane.  With a steep stellar density
distribution, a cloud does not need to be far in physical distance from
the midplane to be outside of 70 percent of the light.  

\section {Globular Clusters} 

We identify 20 bright knots apparent in both bands around NGC~1316 as
globular clusters.  One much brighter object has been shown to be a
Galactic foreground star (Schweizer 1980).   Cluster locations are given
in Figure~\ref{cloc}.  In Table~2 we also report the coordinate
locations, apparent brightnesses, and $V-I$ colors.  Brightnesses were
determined by aperture photometry with a 3 pixel radius and an aperture
correction, based on a star in the field, measured in the same manner.
Since the clusters are extended the total brightnesses will be
somewhat greater. Only the extreme bright end of the globular cluster
luminosity distribution is detected.   

For the purpose of determining if the clusters are resolved, we measured
the fraction of encircled energy at the 3 pixel radii relative to the  6
pixel radii for each cluster and compared these to the same quantity for
the Galactic foreground star.  This can be written:

\begin{equation}
\Delta \equiv 2.5 \log \Bigl({F(6)F^*(3) \over F(3)F^*(6)}\Bigr)
\end{equation}

Figure~\ref{clustercol} presents the $\Delta$ values for all the globular
clusters in Table~2 plotted against absolute magnitude.  In nearly
all cases the value of $\Delta$ is positive, indicating
larger measurable extent than the star, but the error bars, taken from
the DAOPHOT photometry package, are large.  A weighted average value for
 $\Delta$ is 0.075 mag (0.06 for V~band and 0.09 for the I~band). 

To determine if this is consistent with typical globular cluster light
distributions at the distance of NGC~1316, we created simulated images
of self-consistent King models (1966) with a range of core radii and
tidal radii $r_t$ = 25~pc, 50~pc, and 75~pc.  The values of $\Delta$ for
these models are shown in Figure~\ref{king}.  The solid horizontal line
is placed at the measured mean value for $\Delta$ of 0.075.  It appears
that we do not measure core radii for any of the clusters, but we can
put limits on the tidal radii.  A finite tidal radius is required by 
the measured value of $\Delta$ because an infinite tidal radius gives a
larger value at all core radius values.  It appears that the tidal radii
in the region examined are typically $<$ 35~pc.  For comparison, the
tidal radii of the four M31 globular clusters studied by Grillmair {\it
et al.} (1996) range from 35 -- 60 pc. For Galactic globular clusters,
35 pc is in the middle of the range of tidal radii (Trager {\it et al.}
1993;  Djorgovski 1993).

The mean $V-I$ color of the globular clusters is about 0.3 mag bluer
than the galaxy itself within the WFPC field, but that is still 0.2 mag
redder than the mean of Galactic globular clusters and 0.1 to 0.3 mag
redder than the ten globular cluster systems in Virgo measured by  Ajhar
{\it et al.} (1994).  The brightest globular cluster, with $M_V=-12.7$,
is the most luminous red globular cluster to be resolved that the
authors are aware of, although, in other galaxies, there are brighter
candidate clusters that have not been confirmed to be non-stellar.


Couture, Harris \& Allwright (1991) present BVI CCD photometry of 230
globular clusters in NGC~4472. Located in the Virgo cluster, NGC 4472 is
a giant elliptical galaxy at a similar distance. In
Figure~\ref{compare}, we compare the two globular cluster systems. The
distribution of globular cluster colors are similar.  The distribution
of brightest magnitudes in NGC~1316 is also not remarkable, provided
that the total population is comparable in numbers to that in NGC~4472. 
Apparently, although NGC~1316 shows all the signs of a recent merger
with gas rich companion(s), the young, bright, blue globular clusters
seen in other merger events, such as NGC~1275 (Holtzman {\it et al.}
1992) are not seen in NGC~1316.  Either they did not form, or perhaps,
they did form but have since evaporated.

\section {Conclusions} 

Although there is extinction near the line-of-sight to the center of
NGC~1316, it is sufficiently weak at I~band ($A_I < 0.2$ mag) to permit
accurate profile determination.  The central surface brightness is
unusually bright, for a galaxy of this magnitude, and has a very small
projected core radius of $0.41\arcsec$ = 34 pc.   The shape of the
luminosity profile is distinctly nonisothermal.  The deprojected density
reaches 2000 $L_{\sun}~{\rm pc}^{-3}$ in the central 0\farcs05.  The
predicted velocity dispersion calculated from the deprojected density
profile and assuming constant mass-to-light ratio rises with radius. 
However, the observed velocity dispersion is roughly constant (allowing
for the increasing rotation with radius).  To reconcile the observed
dispersion with the brightness distribution requires either a single
dark object of mass $\approx 2 \times 10^9 M_{\sun}$ at the center, a
strongly varying stellar mass-to-light ratio in the inner 100 pc, or a
radially varying velocity anisotropy.  There has been insufficient time
for the steep inner slope or a black hole to have evolved through
stellar dynamical interactions from a larger initial central core
distribution.  The V~band mass-to-light ratio determined outside of 100
pc is $2.2 \pm 0.2$.

Away from the dust lane, NGC~1316 has a fairly uniform $V-I$ color.  The
uniform, very red color precludes a significant young cluster from
explaining the brightness and cuspiness of the inner region. The maximum
extinction regions have A$_V$ of only $\sim 1.5$.  Most of the deep dust
regions detected are found to be in front of at least 80\% of the light
along the line of sight.  This is not very surprising because within
10\arcsec\ of the center, the position in front of 80\% of the light
occurs within 100 pc of the midplane.  It is difficult, in these images,
to discern dust past the midplane from its absorption properties.

Most of the globular clusters associated with NGC~1316 are barely
resolved.  Overall, they appear to be normal in luminosity and color, if
the distance to NGC~1316 is  $\approx 17$ Mpc and the total population
is reasonably large. The one exception is the brightest globular which
has $M_V=-12.7$.  This is more than a magnitude brighter than the
previous brightest resolved globular cluster known. The tidal radii of
the globular clusters seen in the inner 60\arcsec\ appear to be less
than or equal to  35~pc.   There are no bright young globular cluster 
despite the clear evidence of a recent significant merger.  Either
globular clusters were not formed during the merger or they have since
evaporated in the strong tidal field at the center.
                          
\acknowledgments

We wish to thank F. Schweizer for several informative discussions on
this galaxy. This research was conducted by the WF/PC Investigation
Definition Team, supported in part by NASA Grant No. NAS5-1661.

\clearpage

\begin{figure}                        

\caption{Deconvolved Images of N~1316. The whole PC2 chip of WFPC-1 are
shown after pairs of exposures in each filter have been cleaned of
cosmic rays and coadded.  Insert at the lower right are $3\times$
blowups of the central 64 x 64 pixels.  The logarithm of the images is
displayed and the stretch ranges from: a)  40 to 3162 ADU in the F555W
image  (251 to 10000 ADU for the insert) and b) from  50 to 3162 ADU in
the F785LP image (398 to 10000 ADU for the insert). A few of the 
globular clusters can be seen in this reproduction.} \label{fig1}

\caption{Color Map of NGC~1316. $V-I$ image after 60 iterations of the
Lucy-Richardson deconvolution procedure were applied to the V and I
images. All 4 PC chips are shown.  Orientation is the same as Figure 1. 
The dust lane is clearly visible with many interesting filaments and
swirls. }\label{bigvmi}

\caption{Inner region color map. Grayscale image of $V-I$ for just the
central 128 by 128 pixels.  Pixels have been rebinned 2 by 2. Contours of
$V-I$ are superimposed.  The bluest point near the middle is the
central pixel.  Orientation is the same as Figure 1.}\label{vmicr} 

\caption{Brightness profiles as a function of semimajor axis. Plus signs
refer to both the pre-deconvolved and to 60 iterations of
Lucy-Richardson deconvolved I~band images and diamonds refer to the same
for the V~band image. Pixels with $V-I > 1.6$ were not included in the
ellipse fitting procedure. The line on the deconvolved I~band data
is the fit using Equation 1.  The line on the deconvolved V~band data
is the I~band fit plus 1.55.}\label{magprof}

\caption{Deprojected luminosity density profile.  Abel inversion of the 
fit to the deconvolved I~band profile, converted to solar
luminosities pc$^{-3}$.  Solid line is the result of assuming that
the galaxy is an oblate spheroid (i.e., inversion of major axis), and
dashed line is the result of assuming galaxy is a prolate spheroid
(i.e., inversion of minor axis).}\label{lumden}

\end{figure}
\begin{figure}

\caption{Radial velocity dispersion at each deprojected radius. 
Hydrostatic equilibrium was assumed, and a mass-to-light of 2.2 was
applied to the luminosity density profile (previous figure).  The
observed rotation curve of  Bosma {\it et al.} (1985) and the
ellipticity measures of the profile fitting were used to determine the
weight of each oblate shell. Solid line refers to simple single
mass-to-light ratio.  Dashed line is obtained if no rotation is
assumed.  Dotted lines result from adding a compact dark object of
either $3 \times 10^8, 1 \times 10^9$, or $3 \times 10^9
M_{\sun}$.}\label{vel_r}

\caption{Velocity dispersion along a slit.  The radial velocity
dispersions (squared) from Figure 6 were integrated along the line of
sight and weighted by the local densities assuming $\sigma$ is isotropic. 
Line styles are the same as the previous figure.  The velocity measurements
of Bosma {\it et al.} (1985) along $pa=60\arcdeg$ are  plotted with
their error bars.}\label{vels_R} 

\caption{Velocity dispersion within a circular aperture.  The profiles
of velocity dispersions squared were integrated along the line
of sight and weighted by the local densities and then averaged over a
circular aperture of radius R$_{circ}$.  Line styles are the same as Figure~6.
The central velocity measurements of Bosma {\it et al.} (1985) (plus
sign), Schweizer (1981) (triangle) and Jenkins \& Scheuer (1980) (square)
are  plotted with their error bars.}\label{vela_R}

\caption{Mass-to-light ratio variation.   Velocity measurements indicate
a nearly isothermal velocity profile.  This does not agree with the
brightness profile.  The curves here give several possible runs of
mass-to-light with radius that would result in isothermal distributions. 
The numbers next to each curve give the core radii in parsecs of the
corresponding isothermal sphere.}\label{emol} 

\caption{Extinction Map.  A fit to the pixels with little or no
reddening is subtracted from the inner 256 by 256 pixel subimage of the
deconvolved V~band image.  The resulting image vividly portrays the dust
distribution by the amount of light removed from the line of
sight.  The figure gives reassurance that away from dusty regions
the model fit is reasonably good.  Orientation is the same as Figure
1.}\label{v60fitm}

\end{figure}
\begin{figure}

\caption{A cut along the major axis of the V~band fit of \S~5.1 (pluses)
compared to I~band ellipse fitting model of \S~3 (asterisks), offset by
1.55 mag.}\label{v60profs}

\caption{Radio map contours at 1.5 GHz (Geldzahler \& Fomalont 1984) are
overlayed on the $V-I$ color map shown as a grayscale (darker is
redder).  Most of the PC2 chip plus parts of PC1 and PC3 are shown. 
Note the abrupt termination of the radio jet at the absorption features
to the South.  North is up.}\label{radio}

\caption{Scatter plot of $E(V-I)$ versus $V_{obs} \over V_{fit}$ for all
pixels in annulus from 1\arcsec\ to 7\arcsec.  Curves of constant depth
of the position of the cloud, $s_2$, but varying extinction $A_V$ are 
indicated by dashed lines.  Curves of constant $A_V$ but varying depth
of the cloud is indicated by solid lines.  A$_V$ = 2.0 A$_I$ has been
assumed.}\label{avs}

\caption{Scatter plots of models of dust regions after being convolved
with the PSF, noise added to data and PSF, and 60 iterations of
Lucy-Richardson deconvolution.  a) $A_V$ = 1, radius of dust cloud = 10
pixels, depth $s_2$ = 0, b) $A_V$ = 1, radius of dust cloud = 10
pixels, depth $s_2$ = 0.25, c) $A_V$ = 1, radius of dust cloud = 20
pixels, depth $s_2$ = 0, d) $A_V$ = 2, radius of dust cloud = 10
pixels, depth $s_2$ = 0,}\label{avmodels}

\caption{Cluster Locations relative to the center of NGC~1316. We have
included contours from  V~band for reference.  Cluster locations are
indicated by '*'-sign.  Orientation is the same as Figure~1.} \label{cloc}   

\caption{Globular cluster deviation of aperture photometry from
expectations for a point source.  Fraction of light within an aperture
of 3 pixel radius relative to 6 pixel radius for globular clusters
divided by same quantity for a star in the image, $\Delta$ in the text. 
The amplitude is displayed in astronomical magnitudes.  Top figure is
for V~band and bottom figure is for I~band.  The brightest cluster in
the I~band is excluded because it fell on the CCD chip boundary, of
course. Horizontal lines are drawn at the weighted average for each
filter.} \label{clustercol}

\end{figure}
\begin{figure}

\caption{Degree of resolution for simulated King (1966) isothermal
models, scaled to the NGC~1316 distance, as a function of the core
radius of the model. The tidal radii used were  25~pc (dashed), 50~pc
(dotted), and 75~pc (solid).  Solid horizontal line is placed at
measured mean value for the globular clusters.}\label{king}  

\caption{Comparison between clusters of NGC~1316 and NGC~4472. The
histogram of number of globular clusters at each 0.2 mag bin in $V-I$ 
of a) NGC~1316 and b) NGC 4472.  Scatter plot of $V-I$ vs. $V$ for
clusters in c) NGC~1316 and d) NGC 4472.} \label{compare}

\end{figure}

\clearpage
\begin{references}

\reference{} Ajhar, E.A., Blakeslee, J.P., \& Tonry, J.L. 1994, \aj,
108, 2087

\reference{} Bessell, M.S. 1979, \pasp, 91, 589 

\reference{}  Binney, J.  \& Mamon, G. A. 1982, \mnras, 200, 361

\reference{}  Binney, J.  \& Tremaine, S. 1987,  {\it Galactic Dynamics}
(Princeton: Princeton University Press)

\reference{} Bosma,  A., Smith, R.M., \&  Wellington, K.J.,1985,
 \mnras, 212, 301 

\reference{} Bracewell, R. N. 1978, {\it ``The Fourier Transform and Its
Applications"} (New York:  McGraw-Hill) 

\reference{} Byun, Y.I.,  Grillmair, C.J., Faber, S.M., Ajhar, E.A.,
	Dressler, A., Kormendy, J., Lauer, T.R., Richstone, D. 1996, 
	\aj, in press
                   
\reference{} Cohn, H. 1980, \apj, 242, 765

\reference{} Couture, J., Harris, W.E., \& Allwright, J.W.B. 1991,
\apj, 372, 97

\reference{} Djorgovski, S., 1993, in ``Structure and Dynamics of
Globular Clusters: ASPCS vol. 50", S. Djorgovski and G. Meylan (eds.), p
373

\reference{} Fabbiano, G., Fassnacht, C., \& Trinchieri, G. 1995, \apj, 434 67.

\reference{} Faber, S.M., \& Jackson, R. E. 1976, \apj, 204, 668

\reference{} Faber, S.M., Wegner, G., Burstein, D.,Davies, R. L., Dressler,
A.,Lynden-Bell, D., \& Terlevich, R.J. 1989, \apjs {\bf 69}, 763

\reference{} Freedman W. {\it et al.}, \apj, in press

\reference{} Geldzahler, B.J., \& Fomalont, E.B. 1978, \aj, 83, 1047 

\reference{} Geldzahler, B.J., \& Fomalont, E.B. 1984, \aj, 89, 1650 

\reference{} Grillmair, C.J., Ajhar, E.A., Faber, S.M., Lauer, T., Baum,
W.A., \& Holtzman, J.A., \aj, in press

\reference{} Inagaki, S., \& Saslaw, W.C., 1985, \apj, 292, 339

\reference{} Holtzman, J.A., Faber, S.M., Shaya, E.J., Lauer, T.R., Groth,
     E.J., Hunter, D.A., Baum, W.A., Ewald, S.P., Hester, J.
     J.,  Light, R.M., Lynds, C.R., O'Neil, E.J., \& Westphal,
     J. A.  1992,   \aj,  103, 691

\reference{} Holtzman {\it et al.}, 1996, \aj, in press

\reference{} Jenkins, C.R. \& Scheuer, P.A.G. 1980, \mnras, 192, 595


\reference{} King, I. 1966, \aj, 71, 64

\reference{} Kormendy, J. 1987, in IAU Symposium 127: Structure and Dynamics of
Elliptical Galaxies, ed  T. de Zeeuw. (Dordrecht: Reidel), p. 17

\reference{} Kormendy, J, \& Richstone D. 1995, preprint

\reference{} Lauer, T.R. {\it et al.}, 1996, \aj, in press

\reference{} Lauer, T.R., 1989, PASP, 101, 445

\reference{} Lauer, T.R. {\it et al.} 1992, \aj, 104, 522  

\reference{} Lucy, L.B. 1974, \aj, 79, 745

\reference{}  McMillan, R., Ciardullo, R., \& Jacoby, G.H. 1993, \apj,
416, 62

\reference{} Poulain, P. 1988, \aaps, 72, 215



\reference{} Richstone, D.O., \& Tremaine, S. 1986, \aj, 92, 72 

\reference{} Sargent, W.L.W., Young, P.J., Boksenberg, A.,
Shortridge, K., Lynds, C.R., \& Hartwick, F.D.A. 1978, \apj, 221, 731

\reference{} Schweizer, F. 1980, \apj, 237, 303 

\reference{} Schweizer, F. 1981, \apj, 246, 722 


\reference{} Shaya {\it et al.} 1994, \aj, 107, 1675


\reference{} Tonry, J.L. 1991, \apjl, 373, L1

\reference{} Trager, S., Djorgovski, S., \& King, I.R. 1993, in
``Structure and Dynamics of Globular Clusters: ASPCS vol. 50", S.
Djorgovski and G. Meylan (eds.), p 347. 

\reference{} Sandage, A., Tammann, G.A. 1987, {\it A Revised Shapley-Ames
Catalog of Bright Galaxies} (Washington, DC: Carnegie Institution)

\reference{} Tully, R.B., Shaya, E.J., \& Pierce, M.J. 1992, \apjs, 80, 479

\reference{} WF/PC Investigation Definition Team 1992, Final Orbital
Science Verification Report, a Report to the  Space Telescope Science
Institute, Baltimore, Maryland

\reference{} Whitmore, B., Schweizer, F., \& Leitherer, C. 1993, \aj, 106.
1354

\reference{} Whitmore,  B., \& Schweizer, F. 1995, \aj, 109, 960

\reference{} van der Marel, R.P. 1991 \mnras, 253 710

%
%
%
%
%
%
%
%
%
%
%
\end{references}
\end{document}